# Hydrogenated amorphous silicon detectors for particle detection, beam flux monitoring and dosimetry in high-dose radiation environment

**M. Menichelli**[a], **M. Boscardin**[b,c], **M. Crivellari**[c], **J. Davis**[f], **S. Dunand**[i], **L. Fanò**[a,d], **A. Morozzi**[a,g], **F. Moscatelli**[a,e], **M. Movileanu-Ionica**[a], **D. Passeri**[a,g], **M. Petasecca**[f], **M. Piccini**[a], **A. Rossi**[a,d], **A. Scorzoni**[a,g], **L. Servoli**[a], **G. Verzellesi**[b,h] **and N. Wyrsch**[i]

[a] *INFN, Sez. di Perugia,*
  *Perugia (ITALY)*
[b] *INFN TIPFA,*
  *Trento (ITALY)*
[c] *Fondazione Bruno Kessler,*
  *Trento (ITALY)*
[d] *Dip. Di Fisica dell'Università degli studi di Perugia,*
  *Perugia (ITALY)*
[e] *CNR-IOM,*
  *Perugia (ITALY)*
[f] *Centre for Medical Radiation Physics, University of Wollongong,*
  *NSW 2522, (AUSTRALIA)*
[g] *Dip. di Ingegneria dell'Università degli studi di Perugia,*
  *Perugia (ITALY)*
[h] *Università di Modena e Reggio Emilia,*
  *(ITALY)*
[i] *Ecole Polytechnique Fédérale de Lausanne (EPFL), Institute of Microengineering (IMT),*
  *Neuchatel, (SWITZERLAND)*
  *E-mail*: mauro.menichelli@pg.infn.it

ABSTRACT: Hydrogenated amorphous silicon (a-Si:H) has remarkable radiation resistance properties and can be deposited on a lot of different substrates. A-Si:H based particle detectors have been built since mid 1980s as planar p-i-n or Schottky diode structures; the thickness of these detectors ranged from 1 to 50 μm. However MIP detection using planar structures has always been problematic due to the poor S/N ratio related to the high leakage current at high depletion voltage and the low charge collection efficiency. The usage of 3D detector architecture can be beneficial for the possibility to reduce inter-electrode distance and increase the thickness of the detector for larger charge generation compared to planar structures. Such a detector can be used for future hadron colliders for its radiation resistance and also for X-ray imaging. Furthermore the possibility of a-Si:H deposition on flexible materials (like kapton) can be exploited to build flexible and thin beam flux measurement detectors and x-ray dosimeters.

**Contents**

**1. Introduction**

Hydrogenated amorphous silicon (a-Si:H) is a disordered semiconductor studied since the late 1960s. After the first report about the possibility of doping this material both in p and in n-type [1] the manufacturing of several electronic devices was successfully attempted. The utilization of this material for particle detection is reported since 1986 [2,3] when p-i-n and Schottky diodes were successfully manufactured and tested.

The most important feature of a-Si:H is its remarkable radiation resistance, as reported in ref. [4] where a 32.6 μmthick p-i-n diode was irradiated with 24 GeV protons up to a fluence of $7 \times 10^{15}$ p/cm$^2$ and its leakage current increased only by a factor 2 at $9 \times 10^4$ V/cm electric field and this increase was completely reversed after 24 hours of annealing at 100 °C. Despite the extremely high radiation hardness of this material, the main limit of a-Si:H planar detectors is their poor signal-to-noise ratio for the detection of MIPs that never exceeded the value of 5. The reason for this is the very high depletion voltage needed (fields up to $10^5$ V/cm) that generates a very high leakage current (in the order of 1 μA/cm$^2$ ) and the poor charge collection efficiency (about 50% for a 30 μm thick diode) due to the disordered nature of a-Si:H lattice structure. Another possible drawback of this material is the low mobility of charge carriers (1-10 cm$^2$ V$^{-1}$ s$^{-2}$ for electrons and 0.01 cm$^2$ V$^{-1}$ s$^{-2}$ for holes) that makes the charge collection time of about 10 ns in a 20 μm thick diode [5].

In order to overcome these problems we propose to build this detector using a 3D geometry that allows to keep a relatively small collection distance increasing the thickness up to 100 μm or more in order to increase the total charge generated in the detector by a MIP. The reduction of the distance between the electrodes is a crucial factor for keeping the leakage current low reducing the noise. Another possible way to reduce the noise is the operation at low temperatures. The forthcoming vertex detectors at LHC will operate at temperatures around −25°C, since this detector may have application as an alternative to pixel detectors made of crystalline silicon (c-Si), the operation at −25°C may be a benefit from the point of view of signal to noise ratio but may reduce the mobility for the electrons (the mobility of a-Si:H decreases with temperature as shown in ref. [6]) resulting in a slower signal. Therefore the operation of these detector at various temperature should be carefully studied.

## 2. The construction of a 3D a-Si:H detector.

Hydrogenated amorphous silicon layers are usually obtained by deposition of silane on an existing substrate. Known substrates for deposition are: crystalline silicon, glass (normal glass, fused silica, pyrex as well as many special types of glass), stainless steel foils, aluminum (better if coated with chromium, molybdenum, platinum, palladium etc.), kapton©, chromium plated brass, other polymers like: PEN , PET and PI, silicon oxide, coated ceramic, copper coated PCB, heat resistant organics/inorganic polymers (like ormocer©) on a supporting wafer etc.

For our application we selected heavily doped p-type silicon wafers because in one of the detector configurations we plan to build, it may be useful as a bias plate as we will discuss below. On top of this substrate material we will deposit a-Si:H by PECVD using a VHF excited plasma at the frequency of 70 MHz from a mixture of silane and hydrogen at temperatures around 200°C. Other techniques for depositing device-grade a-Si:H are PECVD at radio-frequency or microwaves, or hot-wire deposition. The resulting material is a disordered semiconductor with a short-range order where not all the Si-Si bonds are actually saturated. The number of dangling bonds is reduced by the introduction of Hydrogen atoms. Best quality materials exhibit 4 to 10% atomic hydrogen content. More details on the deposition of a-Si:H for detector application can be found in Ref. [7]. After a-Si:H deposition on the substrate, silicon nitride will be deposited on the a-Si:H layer for passivation using PECVD at low temperature (e.g. $\leq 250$°C).

In order to fabricate the electrode, holes (or trenches) needs to be etched on the passivated a-Si:H surface. The technique that will be used for this purpose is Deep Reactive Ion Etching (DRIE). With the DRIE process it is possible to fabricate holes with few micron diameter maintaining the process temperature below 250 °C.

Once the holes are etched, in order to build the basic p-i-n electrode structure of the detector, there is the necessity to dope the a-Si:H material in the internal surfaces of the holes. Since commonly used techniques for planar structures (i.e. PECVD deposition of doped a-Si:H) are not applicable for this geometry (deep and narrow holes), two options will be considered:

- Option 1- Atomic Layer Deposition (ALD) of conductive metallic oxides to create selective contacts: titanium oxide could be used for electron selective contact and tungsten or molybdenum oxide for hole selective contacts.
- Option 2- Ion implantation of phosphor for n-type doping and boron for p-type doping and subsequent annealing (if required) at low temperature (e.g. $\leq 250$°C)

Option 1 has been used in solar cell fabrication [8] but never demonstrated for detectors. Important device characteristics for the particular application such as the carrier collection efficiency and the leakage current as a function of applied electric field have never been adequately studied. Option 2 has been demonstrated in ref. [9] where an electric junction based on implanted ions in a-Si:H has been successfully fabricated. Since these processes are not very common, a prototyping phase is foreseen where these two techniques will be used in the construction of planar p-i-n diodes as described below. As the last step of the process chromium will be used as conducting metal for contacts since it can be deposited at low temperature and also because it does not diffuse into a-Si:H; such a material has already been used in detector fabrication. In order to improve soldering capability, aluminum will be deposited over chromium by sputtering or thermal evaporation.

The proposed configurations for a-Si:H 3D detectors are shown in Fig.1. The option a) foresees a grid shaped metal layer for the bias of p-type electrodes and separate contacts for n-type electrodes that collect the signal. These electrodes can be read individually or grouped in order to reduce the number of readout channels. The option b) uses the c-Si substrate (p-type doped) as biasing plane in contact with an Aluminum layer; the readout for the n-type electrode is the same as in option a).

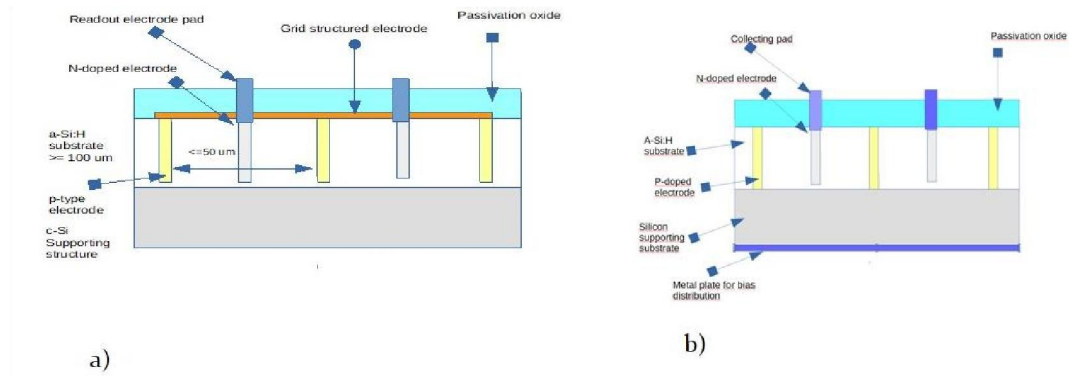

Fig.1. Two configurations for detector vertical layout. In configuration a) the bias voltage is connected to a grid structured metal plane which is in contact with p-type electrodes. In configuration b) the p-type c-Si substrate is used as a contact for the p-type electrodes.

Various geometrical shapes and layouts of electrodes are foreseen, as shown in Fig. 2. Basically according to their geometrical shape the electrodes are of three types: finger-shaped, mini-trench-shaped and full-trench-shaped.

For the purpose of evaluating the various technological options towards the fabrication of a 3D a-Si:H detector we foresee three phases of development:

1. Fabrication of planar detectors in two sets of prototypes: one uses ALD for doping (option 1) while the other one uses Ion implantation (option 2)
2. Fabrication of basic 3D structures (fingers and trenches) in various configurations both with option 1 and option 2 doping technologies.
3. Fabrication of the detectors according to the various configurations in Fig. 2 in both options 1 and 2 doping technologies.

Currently we are in the first phase of prototype development. Planar devices are under fabrication in two main types: vertical diodes and lateral diodes. Vertical diodes use the supporting low resistivity p-doped substrate as biasing electrode and the n-type electrodes are doped usingone of the two doping options. The geometries for these detectors are shown in Fig. 3. Lateral devices have both the electrode on the top side and the supporting c-Si substrate is insulated by silicon oxide (or Silicon Nitride) from the a-Si:H layer. The various electrode configurations for these devices are shown in Fig.4. Both these devices have in common the thickness of the a-Si:H layer which is 10 μm.

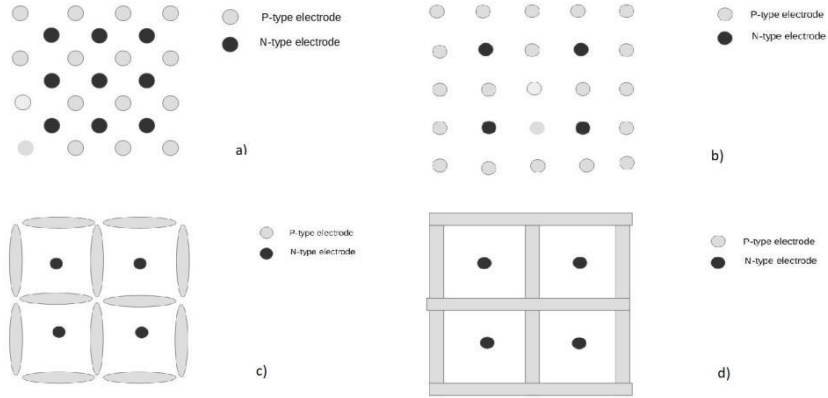

**Fig.2**. Four possible electrode configurations: a) baseline structure with n-type electrode (signal collecting electrode) surrounded by 4 p-type electrodes b) n-type electrode surrounded by 8 p-type electrodes. c) n-type electrodes surrounded by p-type mini-trenches. d) n-type electrodes surrounded by p-type full trenches

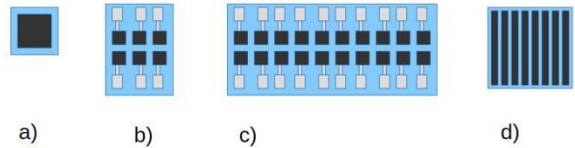

**Fig. 3.** Vertical devices a) Single 2.5 × 2.5 mm² diode. b) 2 × 3 diode array available with or without bonding pad and in two diode sizes 1 × 1 mm² or 0.5 × 0.5 mm² c) 2 × 10 diode array, diode size 1 × 1 mm² available with or without bonding pad. d) 8 strip devices in two sizes: with strip 5 × 0.2 mm² and 10 × 0.5 mm² available with or without bonding pad.

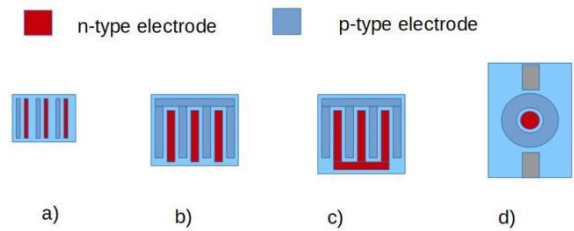

**Fig. 4.** Lateral diodes. a) 5 lateral devices array (only 3 shown) with 10, 20 and 30 μm spacing. Size of each electrode is 0.2 × 5 mm² b) comb like devices with 10, 20 and 30 μm spacing Size of each n-type electrode is 0.2 × 5 mm² c) double comb like devices 10, 20 and 30 μm spacing sizes equal to b). d) dosimetric device with 100 μm central n-type electrode and 10, 20 and 30 μm spacing from the n to the p-type electrode.

## 3. The simulation of a-Si:H material

In order to optimize the design and construction of 3D a-Si:H detectors and the relevant test structures an appropriate simulation should be developed, however at present no established commercial simulation package of a-Si:H that can mimic the electrical and charge collection behavior of this material within the framework of radiation detection is available. Some effort has been performed to simulate thin film devices (i.e. nm thickness range) for photovoltaic applications using a variety of commercial and non-commercial software packages. The first attempt of developing this kind of software was performed by Dutta that developed a numerical code especially devoted for the optimization of the sensitivity and open circuit voltage for a-Si:H based p-i-n solar cells [10]. Subsequently Despeisse developed a HSPICE model that was able to simulate the signal generated by pulsed laser in an a-Si:H particle detector with successful results [11]. The first attempt to use a commercial Technology Computer Aided Design (TCAD) simulation tool-kit (from SILVACO$^©$) in the simulation of a-Si:H solar cells [12] was performed by Nawaz. Nawaz's model is based on the definition of defects as a combination of single energy level defects within the 'mid-gap'. In recent years Technology Computer Aided Design developers moved to a SYNOPSYS$^©$ TCAD named Sentaurus. Unfortunately a-Si:H is not modeled within the standard material libraries of Sentaurus, therefore all a-Si:H users should develop their own libraries. Lee, et al, 2009, was the first to develop such a library in Sentaurus TCAD for the simulation of the performance of an a-Si:H based device [13]. Sentaurus TCAD was also used by Otero et al. in 2011, for the modeling a-Si:H solar cells [14]. The n-i-p structure simulated by Otero although much thinner, is very similar to the one that is investigated in this work. This study considered the defects present within a-Si:H as a continuous exponential tail distribution originating from the conduction and valence bands, and a Gaussian distribution for the mid-gap trap states using the Density of States (DOS model). However these models were more suitable to simulate the behavior of solar cells while in our study, a more particle oriented approach should be performed. The main focus of our model should be electric field, dark current, charge collection efficiency and signal formation within the detector irradiated with MIPs, Ions or X-rays. The resulting numerical model for the a-Si:H material should be used in the optimization of the geometries and doping profiles structure of semiconductor devices for particle detection.

As previously stated Sentaurus TCAD does not include a specific model for the simulation of a-Si:H therefore it is necessary to implement a customized material parameter file that contains the specific values of the various parameters for a-Si:H. The file that is commonly used for silicon (called 'Silicon.par') has been then replaced with this custom file named 'aSiH.par'. The values of the specific parameters included in this file are shown in Table 1 compared with the range of values that can be found in the literature.

| Parameter (units) | Literature | Simulation |
|---|---|---|
| Relative permittivity | 11.8-11.9 | 11.7 |
| Electron mobility (cm$^2$ V$^{-1}$ s$^{-1}$) | 0.5-20 | 10 |
| Hole mobility (cm$^2$ V$^{-1}$ s$^{-1}$) | 0.003-10 | 0.01 |
| Band Gap (eV) | 1.7-1.9 | 1.84 |
| Dangling bond density (cm$^{-3}$) | 5 x 10$^{14}$ – 5 x 10$^{16}$ | 4 x 10$^{15}$ |
| Activation energy (eV) | 0.42-0.49 | 0.52 |

**Table 1** Main a-Si:H values in the parameter file.

Aside from the development of this new material parameter file, a description of the defect distribution should be implemented within the device physics module. In this framework both acceptor (electron trap) and donor (hole trap) defects at discrete energy levels within the mid-gap should be implemented. In order to do this implementation the two level defect model described by Petasecca, et al [15] modified to agree with the defect model proposed by Nawaz was implemented. In order to obtain agreement with the experimental results, unique scaling factors were applied to the concentration of individual acceptor/donor defects in the tails and Gaussian distribution of defects.

In order to mimic properly the leakage current of an a-Si:H device the Poole-Frankel model of conduction has to be included in the simulation. However the module which is embedded in Sentaurus TCAD has been designed for organic semiconductors and after some comparative test with real devices it has been removed from the simulation and a fully customized solution has been implemented. Basically the leakage current has been simulated according to the following semi-empirical formula, where a, $J_0$ and $b_0$ are fitting parameters, d is the thickness of the a-Si:H layer, $E_0$ is the activation energy and E is the electric field.

$$J = J_0 \left( a * \exp^{\sqrt{E}*b_0*\sqrt{d}} + \exp^{-\frac{E_0}{kT}} \right)$$

In order to tune some empirical parameters and to verify the correctness of our simulation we compare the results of our simulation program with the measurements taken from two devices having thickness of 12 and 30 μm. The simulated model features a simplified 2D version of the n-i-p (n-doped, intrinsic and p-doped layers) a-Si:H diode structure as described by Wyrsch, et al [7]. The simulated device, features a 90 nm thick n-type layer upon a 30 or 12 μm thick intrinsic layer upon a 90 nm thick p-type layer. The p-type and n-type doped layers are modeled using a Gaussian analytic function with peak concentrations of 3 × 10$^{18}$ and 8 × 10$^{18}$ cm$^{-3}$, respectively. The modeled device is 50 μm wide and given its 2D nature, is by default 1 μm long. Given that the model is a 2D representation of the a-Si:H n-i-p device, a geometric scaling factor $A_F$ was used to allow for a comparison between experimental and simulated leakage current density. This geometric scaling factor is used to account for the considerably different areas of the experimental (2 × 2 mm$^2$) and simulated (50 × 1 μm$^2$) devices structures.

The application of the leakage current simulation model to the detectors described above has produced results in agreement with the measurements. Both the results for the two detectors of different thickness gave satisfactory results. Also the electric field has been simulated. The simulation shows that the 12μm thick a-Si:H n-i-p structures have depletion for applied bias between 48-96 V, respectively, corresponding to a biasing field strength with a magnitude of 4-8Vμm$^{-1}$. Whilst the DC results of the 30μm thick a-Si:H n-i-p structures show full depletion for applied bias between 300-360 V, respectively, corresponding to an biasing field strength with a magnitude of 10-12Vμm$^{-1}$.

The simulation of charge collection efficiency was performed using the 'HeavyIon' model included in the standard Sentaurus libraries. The charge generation from a MIP was set to 1.28 x 10$^{-5}$ pCμm$^{-1}$ corresponding to approximately 80 pairs/μm. The Charge Collection Efficiency (CCE) is defined as the integral of the current over a time window of 3 ns and is normalized with respect to the maximum CCE observed confirm that the depletion voltage is around 70V for the 12 μm thick diode and about 300 V for the 30 μm thick diode. A complete review of the simulation methods and results can be found in ref. [16].

In conclusion the agreement with the data is quite good and this allows us to refine this simulation work also with the application of the model to the detector prototypes which are under construction.

## 4. A-Si:H detectors for dosimetry and beam flux monitoring

The a-Si:H has some advantageous features that could be exploited for dosimetry and beam monitoring applications. The most important are the insensitivity to radiation damage, the possibility to create very thin active layers with very small area (1 mm$^2$ or less), the small charge collection efficiency that will allow an increase of the dose-rate measurable range, the fast response to an external stimulus.

For these applications, since single particle detection is not required, the less complex planar geometry for a detector thickness in the order of 10-20 μm is preferred. The possibility of deposition in a large number of substrates on a wide area could also be used to create a system of diodes that could measure the radiation geometrical shape at the same time.

The preferred substrate for these applications will be kapton$^©$ or any other flexible polyamide material. If the substrate is also very thin, the possibility of a wide area transmission detector with many small pixels could be explored. Another application could be the construction of an active flange that measures beam fluxes inside the beam pipe with little perturbation of the beam itself.

Also the flexibility of the substrate is an important feature, allowing in personal dosimetry for the fabrication of sticky tape dosimeters that can be attached to the human body, and in the beam flux monitoring to wrap an array of detectors around the beam pipe or in any other reasonable 3D shape to detect beam losses.

A possible scheme of a beam monitoring detector is shown in Fig. 5. An array of planar p-i-n detectors of a convenient size and number can be deposited on a 70 μm polyamide sheet. The sheet can be used also as a flexible PCB to carry bias and current signals from the various devices to a readout system.

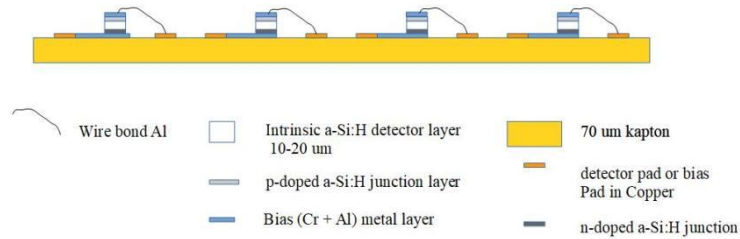

**Fig. 5.** Lateral view (simplified) of an array of diodes on a polyamide PCB.

## 5. Conclusions

3D detectors on Hydrogenated Amorphous Silicon may be a viable option for future colliders where proton fluencies may exceed $10^{16}$ p/cm$^2$ exploiting the very high radiation resistance of the material and if the performances of the 3D detector geometry will be proven as adequate. Thanks to the possibility of depositing a:Si:H on flexible polyamide, applications of this material can also be found in other high radiation environments like in radiation beam flux monitoring and also in personal dosimetry.